\documentclass[prb,twocolumn,showpacs,floatfix,amsmath]{revtex4-1}

\usepackage{color}
\usepackage{dsfont}
\usepackage[normalem]{ulem}
\usepackage{dcolumn}

\usepackage{graphicx}
\def\be{\begin{equation}}
\def\ee{\end{equation}}
\def\beq{\begin{eqnarray}}
\def\eeq{\end{eqnarray}}

\begin{document}
\title{RKKY-like contributions to the magnetic anisotropy energy: 3d adatoms on Pt(111) surface}
\author{Mohammmed Bouhassoune}\email{m.bouhassoune@fz-juelich.de}
\author{Manuel dos Santos Dias}
\author{Bernd Zimmermann}
\author{Peter H. Dederichs}
\author{Samir Lounis}\email{s.lounis@fz-juelich.de}
\affiliation{Peter Gr\"unberg Institut \& Institute for Advanced Simulation, 
Forschungszentrum J\"ulich \& JARA, D-52425 J\"ulich,
Germany}

\begin{abstract}
The magnetic anisotropy energy defines the energy barrier that stabilizes a magnetic moment. 
Utilizing density functional theory based simulations and analytical formulations, we establish that 
this barrier is strongly modified by long-range contributions very similar to Friedel oscillations 
and Rudermann-Kittel-Kasuya-Yosida interactions. Thus, oscillations are expected and observed, with different decaying 
factors and highly anisotropic in realistic materials, which can switch non-trivially the sign of the magnetic anisotropy energy. 
This behavior is general and for illustration we address transition metals adatoms, Cr, Mn, Fe and Co deposited on Pt(111) surface. 
{We explain in particular the 
mechanisms leading to the strong site-dependence of the magnetic anisotropy energy observed for Fe adatoms on Pt(111) surface as revealed previously via first-principles 
based simulations 
and inelastic scanning tunneling spectroscopy (A. A. Khajetoorians et al. Phys. Rev. Lett. {\bf{111}}, 157204 (2013)). The same mechanisms are 
probably active for the site-dependence of the magnetic anisotropy energy obtained for Fe adatoms on Pd or Rh(111) surfaces and for Co adatoms on Rh(111) surface 
(P. Blonski et al. Phys. Rev. B {\bf81}, 104426 (2010)). } 
\end{abstract}
\maketitle




\section{Introduction}
As magnetic devices shrink toward atomic dimensions with 
the ultimate goal of encoding information in the smallest possible magnetic entity, 
the understanding of magnetic stability down to the single atomic limit becomes crucial. Here,  
a critical ingredient is the magnetic anisotropy energy (MAE) that provides directionality and stability to magnetization. 
The larger the MAE, the more protected is the magnetic bit against, for example, thermal fluctuations.  
Thus the search for nanosystems with enhanced MAE is a very active field giving the perspective of stabilizing and simultaneously reducing the size of magnetic bits. 

Recently, it was demonstrated that nanostructures with only a few atomic spins, ranging from single atoms, clusters on metal 
surfaces (see for example Refs.~\cite{Gambardella,Bode,Rau,Khajetoorians_PRL,Dubout,Honolka,Khajetoorians_Science,Otte,Krause,Gambardella2} 
to molecular magnets (e.g. Refs.~\cite{Sessoli,Gatteschi,Brede,Lodi_Rizzini}) can exhibit MAEs that are large enough 
to maintain in principle a stable spin orientation at low temperatures.
A celebrated example is the giant MAE ($\sim$ 9 meV) discovered by Gambardella et al.~\cite{Gambardella} for a single Co adatom 
on Pt(111) surface. There the right ingredients for a large MAE are satisfied: a large magnetic moment carried by the 3$d$ transition element, Co, being 
at the vicinity of heavy substrate atoms characterized by a large spin-orbit interaction (SOI). Naturally, here details of the electronic structure and 
hybridization effects are decisive. Thus exchanging the Co adatom by an Fe adatom leads to an extremely small MAE as demonstrated recently by 
inelastic scanning tunneling spectroscopy and ab-initio simulations based on density functional theory (DFT)~\cite{Khajetoorians}. Most intriguing in the latter work is the dramatic change 
of the MAE magnitude and sign once the Fe adatom was moved from an \textit{fcc}--stacking site, where the moment points out-of-plane, to an \textit{hcp}--stacking site, 
where the moment lies in-plane. This was assigned to the proximity effect leading to a large spin polarization cloud induced by Fe in the Pt substrate, which is notorious 
for its high magnetic polarizability~\cite{Sipr,Meier} as seen also for Pd~\cite{Nieuwenhuys,Herrmannsdorfer,Oswald,Swieca,Mitani}. 
A similar site-dependent MAE for Fe adatoms on the (111) surfaces 
of Pd and Rh and for Co on Rh(111) was noticed with ab-initio simulations~\cite{Blonski}. The physical 
mechanism behind such a behavior has not been, to our knowledge, identified convincingly.  
Even on surfaces with low polarizability, such as gold, 
the MAE follows an oscillating behavior depending on the distance to the surface of buried magnetic nanostructures~\cite{Szunyogh,Szunyogh2,Aas}. 
Thus the polarizability is probably not the only ingredient modifying the strength of the MAE since Au is much less polarizable than Pt.
 One has to keep in mind that the polarizability of the substrate atoms is determined by the Stoner 
product $I \cdot N_F$ with the exchange integral $I$ and the number of states at the Fermi level $N_F$ 
($I \cdot N_F = 0.29$ for Ir, 0.59 for Pt and 0.05 for Au)~\cite{Sigalas}.

The goal of our work is to demonstrate with a formal proof that a strong contribution to the MAE can be highly non-local and long-ranged and may endow up to 
$\pm 50 \%$ of the total MAE. Strong similarities can be foreseen with respect to Friedel~\cite{Friedel} and 
Rudermann-Kittel-Kasuya-Yosida (RKKY)~\cite{RKKY} oscillations in terms of the impact of the 
nature of the mediating electronic states, their localization in real-space and their shape in reciprocal-space (e.g. Fermi surface) on the decay of the oscillations 
and their focusing ( see e.g. Refs.~\cite{Crommie1,Weismann,Lounis,Avotina,Bouhassoune,Avouris1,Zhou,Khajetoorians2,Pruser,Lounis_PRL}). A particularity of this long-range 
contribution to the MAE is, as expected, its dependence on the strength of SOI. Taking as an illustration 3$d$ adatoms (Cr, Mn, Fe and Co) deposited on Pt(111) surface, 
we demonstrate that the contribution of the substrate Pt 
atoms to the total MAE oscillates and decays with their distance to the adatom.

\section{Method}
The MAE can be determined from the magnetic force theorem~\cite{Mackintosh,Jansen} taking the energy difference, 
$\epsilon_{\perp} - \epsilon_{||}$, between 
the band energies of the two configurations: out-of-plane ($\perp$) and in-plane ($||$) orientations of 
the magnetic moment. A reference magnetic configuration is chosen, here the out-of-plane orientation, where the self-consistent calculations 
are performed and the related band energy is obtained. Then the magnetic moment is rotated in-plane and one iteration is done in order to extract the band energy. 
 With such a traditional technique, one reduces the error made by taking differences between the total energies, which are large numbers. 
 A positive sign of the MAE indicates an in-plane preferable orientation of the adatom's magnetic moment.
We utilize the full potential relativistic Korringa-Kohn-Rostoker Green function 
method (FP-KKR-GF)~\cite{Papanikolaou,Bauer}. The local Spin Density Approximation as parametrized by Vosko, Wilk and Nusair was used~\cite{LDA}.
First, the electronic structure of a 22 layers Pt slab with two additional vacuum regions (8 layers) 
is calculated. The experimental lattice parameter (3.92 \AA) was considered without surface relaxations 
which are negligible~\cite{Blonski2}. 
Then, each adatom is embedded on the surface of 
this slab, in real space, together with its neighboring sites, defining a cluster of atoms, where the charge is allowed to be updated during 
self-consistency. We note that the cluster still interacts with the rest of the host surface via the Coulomb interaction. The adatoms 
are allowed to relax towards the surface, and we found qualitatively a similar behavior for the magnetic moments and the MAE in the range of relaxation 
from 15 to 25\% towards the surface. As indicated in Ref.\cite{Khajetoorians}, Fe was found to relax by 20\% towards the surface. The same relaxed geometry was found for 
Co adatoms\cite{Schweflinghaus2016}. Thus for the sake of comparison, the four investigated adatoms were assumed at the same relaxed position 20\% towards the surface.  
The MAE is extracted for clusters of different sizes, for which the Green functions of the impurity-free surface are generated with 
 $200\times200$ k-points in the two-dimensional Brillouin zone and a maximum angular quantum number $l = 3$. 
To provide an idea of the convergence of the MAE versus the number of k-points we address the case of the 
Fe adatom in contact with the Pt substrate where 221 Pt atoms are allowed to be perturbed by the impurity. The MAE is found to 
change by about 0.002\%  with respect to one obtained for $200\times200$ k-points when the number of k-points is decreased to $180\times180$ or $150\times150$ k-points.

\section{Results and discussions}
\subsection{Fe adatoms, {\it fcc} versus {\it hcp} stacking sites}
\begin{figure}[ht!]
\begin{center}
\includegraphics*[angle=0,width=1.\linewidth]{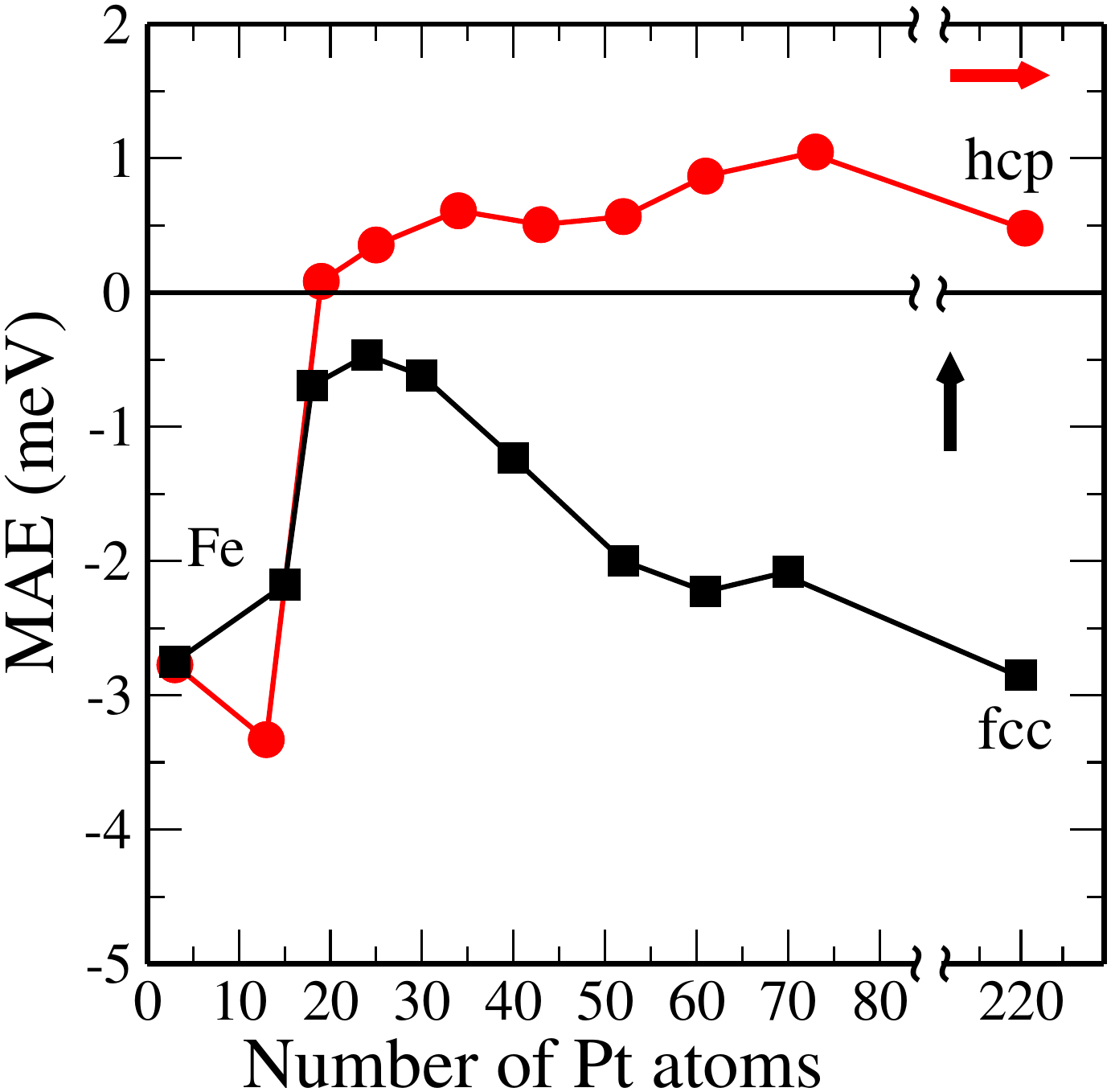}
\end{center}
\caption{MAE of Fe impurity adsorbed on an  \textit{fcc} (square) or an \textit{hcp} (circle) site on top of Pt(111) surface versus the number of Pt atoms in the cluster. A 
positive MAE corresponds to an in-plane orientation of the magnetic moment and a negative MAE corresponds to an out-of-plane magnetic moment. Readapted from 
Fig.3 of the Supplement of Ref.~\cite{Khajetoorians}.}
\label{MAE_1}
\end{figure}

\begin{figure}[ht!]
\begin{center}
\includegraphics*[angle=-90,width=1.\linewidth]{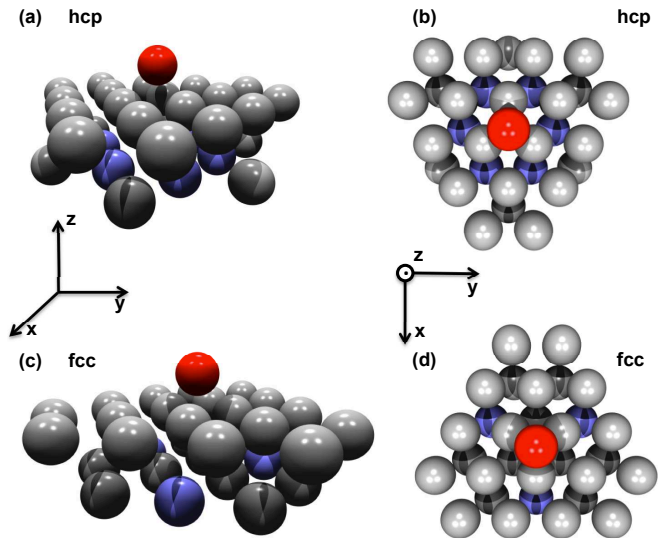}
\end{center}
\caption{Atomic structures of Fe impurity adsorbed on an \textit{hcp} a (side view) and b (top view) or an \textit{fcc} site c (side view) and d (top view) of Pt(111) surface. 
The Pt atoms with blue color are the switching atoms  that have a large contribution to the MAE.}
\label{MAE_2}
\end{figure}

Fig.~\ref{MAE_1} displays the MAE obtained with the band energy differences of an Fe adatom sitting on an \textit{fcc}- or an \textit{hcp}-site on Pt(111) surface 
versus the number of Pt atoms included in the real-space calculations. This figure is part of the Supplement of Ref.~\cite{Khajetoorians}. If only the nearest 
neighbors (NN) Pt atoms to the Fe impurity are considered, in this case 3 Pt atoms, the MAE yields an out-of-plane 
easy axis with the same value of -2.8 meV for both binding sites. However, considering more Pt atoms, the neighborhoods of 
the two stacking sites differ, and therefore the MAE becomes 
strongly dependent on the binding site and even changes sign for the \textit{hcp}-site. The latter occurs when including in the surrounding 
cluster 16 Pt atoms in addition to the NN atoms. The MAE first decreases from -2.8 meV to -3.3 meV by 
adding the 10 closest Pt neighboring atoms and then surprisingly jumps to +0.1 meV by adding the further distant 6 Pt atoms (colored in blue in Fig.~\ref{MAE_2}a-b). 
The latter means that these 6 Pt atoms, with their positive contribution (+3.4 meV) to the MAE, play a key-role in 
switching the preferable orientation of the adatom's magnetic moment.  
These {\it switcher} atoms are equivalent, belong to the subsurface-layer and are equidistant ($\sim 0.5$ nm) 
from the adatom. Interestingly, the switcher atoms occur also for the \textit{fcc} binding site, and are located similarly to the \textit{hcp} binding site 
at the subsurface-layer equidistantly from the adatom. However, their number is lower than in the \textit{hcp} stacking site: 3 instead of 6 (see Fig.~\ref{MAE_2}c-d). 
Therefore, their contribution to the MAE (+1.5 meV) is about half their contribution for the \textit{hcp} binding site. This is not sufficient to compete 
against the preferable orientation of the adatom and its NN (MAE = -2.8 meV). Indeed, once the switching atoms included 
the MAE jumps from -2.2 meV, obtained 
with a cluster containing 15 Pt atoms, to -0.7 meV.  After adding more substrate atoms, reaching a cluster 
of $\sim$ 221 atoms, the MAE tends to +0.5 meV and -2.9. meV for the \textit{hcp} and \textit{fcc} binding sites, respectively.
 The latter values are rather converged since smaller clusters with a number of atoms (not shown in Fig.~\ref{MAE_1}) close to the largest one show a 
stable MAE.

\begin{figure}[ht!]
\begin{center}
\includegraphics*[angle=-90,width=1.3\linewidth]{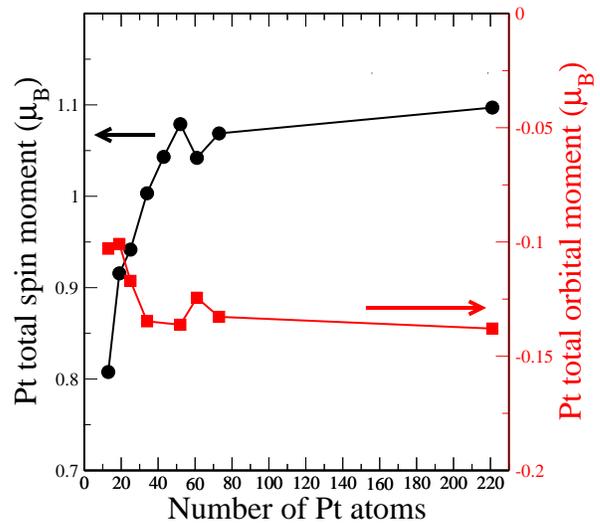}
\end{center}
\caption{Convergence of the total z-component of the spin moment and orbital moment induced in the Pt atoms of different cluster sizes. The case of an Fe adatom sitting on the 
\textit{hcp} binding site is considered and the z-direction is perpendicular.}
\label{Total_Spin_orbital}
\end{figure}

We have also examined the effect of the Pt polarization cloud on the total spin and orbital magnetic moments. Interestingly, the impact on the total moment is less 
impressive than on the MAE as summarized in the Table 1 for the case of the Fe adatom with a magnetic moment pointing out-of-plane. 
When only the NN Pt atoms are included the total spin moment reaches a value of 
$\sim4\mu_B$ while the total orbital moment is around 0.2$\mu_B$. 
Inclusion of a larger number of neighboring Pt atoms increases the total spin moment by a 
maximum of $\sim 0.4\mu_B$ while the total orbital moment reached saturation already with the NN atoms. This observation can be extracted from Fig.\ref{Total_Spin_orbital} where the induced Pt total z-components of the spin and orbital moments are plotted for the case the impurity sits at the \textit{hcp} stacking-site. The z-direction is perpendicular to the substrate. 

\begin{table}[ht!]
\begin{tabular}{c|cc|cc}
\hline
\multicolumn{1}{c|}{ } & {Fe \text{hcp}}& { } &{Fe \textit{fcc}}&{}\\ 
\hline
 Number of Pt atoms   & $m_s$  &  $m_{orb}$ &   $m_s$ &  $m_{orb}$  \\ 
\hline
  3 &4.11&0.22&4.03&0.23 \\
\hline
 53  &4.57&0.216&4.427&0.227 \\
 \hline
 221  &4.59&0.21&4.42&0.212 \\ 
\hline
\end{tabular}
\caption{Total magnetic spin and orbital moments of the Fe adatom including different sets of neighboring Pt atoms. The total spin moment converges after considering 53 Pt atoms, while the total orbital moment 
is already saturated with the NN Pt atoms.}
\label{table1}
\end{table}

\begin{figure*}[ht!]
\includegraphics*[angle=-90,width=1.\linewidth]{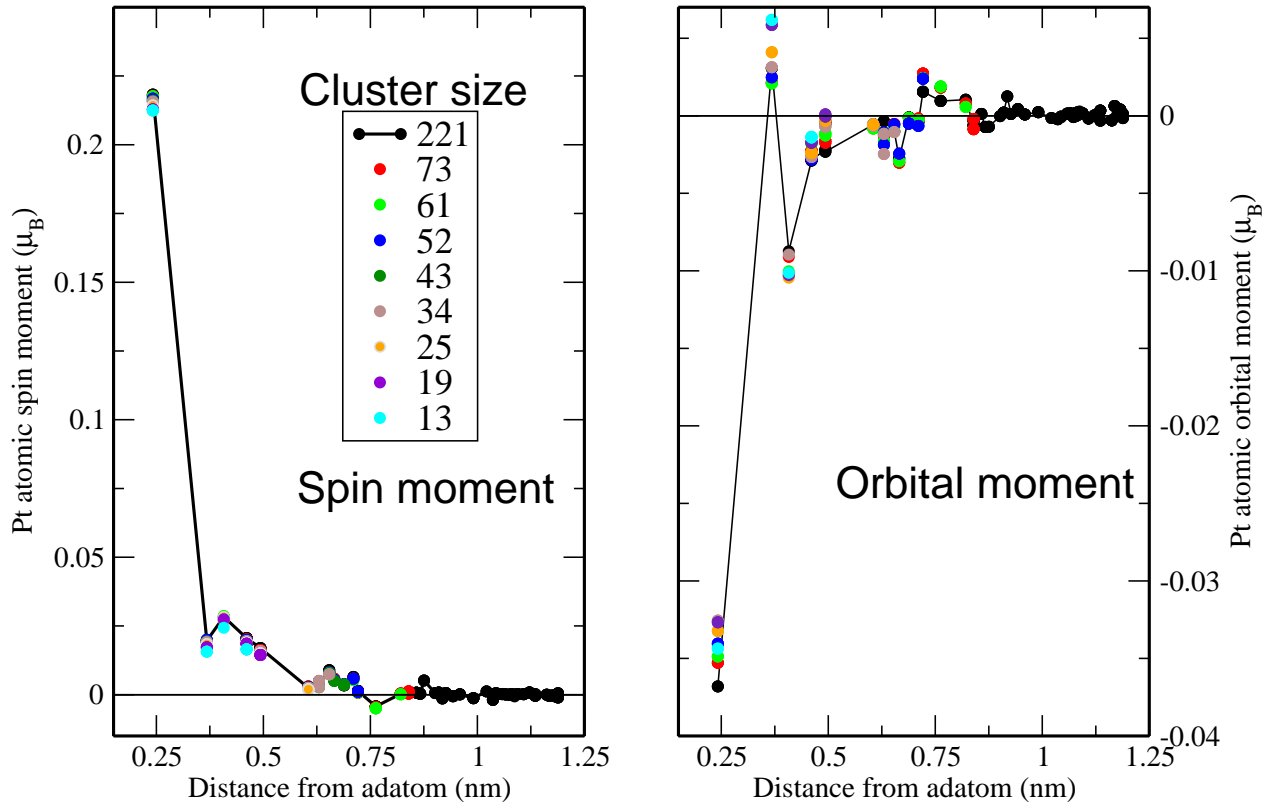}
\caption{The individual Pt atomic spin moment (left) and orbital moment (right) as function of distance with respect to the Fe adatom sitting 
on the \textit{hcp} binding site. The spread of the magnetic moments for the different cluster sizes is rather small, 
highlighting the low impact of 
the boundary conditions of the KKR simulations on these magnetic properties. }
\label{Individual_Spin_orbital}
\end{figure*}

As a summary, one realizes that the contributions of the different Pt shells to the total MAE is not uniform and 
oscillates with the distance and is certainly not correlating perfectly with 
the change of the total spin moment or total orbital moment. The latter quantities describe the polarization of the Pt cloud. 
 At first sight, one could ask whether Fig.~\ref{MAE_1} is the result of numerical artifacts related to the KKR embedding scheme. In principle, 
whenever a cluster is considered, the atoms sitting at the edge of the cluster would feel the boundary conditions more strongly than the atoms close to the Fe impurity. As illustrated 
in Fig.~\ref{Individual_Spin_orbital}, the edge atoms are not that affected by the boundary conditions. 
The spread of the plotted values gives an idea on the impact of the cluster size on the individual Pt magnetic moments. 
As an example, the spin moment of the edge atom, located at $\sim0.65$ nm, in the cluster 
containing 34 atoms is on top of the spin moment of the same Pt atom when the boundary conditions have been improved by extending the size of the cluster to 220 Pt atom. The same 
conclusion can be drawn for the orbital moment, although here the values are much smaller than the spin moments. In general, the boundary conditions will affect slightly the values 
obtained for the magnetic properties including the MAE. However, the general oscillatory behavior observed in Fig.~\ref{MAE_1} seems to go beyond the numerical conditions needed to 
extract it. The main reason is that the Pt spin moment, for instance, has two contributions: either induced by the magnetic adatom or by the surrounding magnetic Pt atoms. 
The former has in general a much stronger contribution than the latter. Also, within the KKR embedding scheme the atoms at the edge feel 
the Coulomb interaction of the neighboring atoms beyond the cluster.

In the following the 
origin of the oscillatory behavior of the MAE will be discussed by realizing that the band energies, $\epsilon$, can be evaluated from $-\int\limits_{-\infty}^{E_{F}}dE\, N(E)$, i.e. an integration up to the Fermi energy, $E_F$, 
of the integrated density of states (IDOS), $N(E)$, which in turn can be extracted from the celebrated Lloyd's formula~\cite{Lloyd}. Indeed, if a system described by 
a Green function, $G$, is perturbed by a potential $V$, the change in the IDOS, $\delta N(E)$, is given simply by $-\frac{1}{\pi}\Im \mathrm{Tr}\,\mathrm{ln}(1 - V G(E))$, where the trace is taken over the site index, 
orbital and spin angular momentum quantum numbers. This permits the aforementioned decomposition of the MAE into local and non-local contributions by evaluating wisely the 
change in the IDOS.

\subsection{Long-range contributions to the MAE: Formalism and results}

First, we note that once the adatom is deposited on the substrate, it 
perturbs strongly the potentials of the NN Pt atoms, which are 3 Pt atoms in total. 
If we consider solely the adatom and its NN Pt atoms, the corresponding Green function, $G_1$, can be obtained from the Dyson equation 
 \begin{equation}
 G_{1}(E)= G_{0}(E)+G_{0}(E)V_{1}G_{1}(E),\
\end{equation}
where $G_0$ is the Green function of the ideal surface of Pt without SOI while $V_1$ is the perturbing potential limited to the region of the adatom and its NN and 
is induced by the presence of the impurity and the SOI. Instead of the potential 
$V_1$, one can use the scattering matrix $T_1$:
\begin{equation}
G_{1}(E)=G_{0}(E)+G_{0}(E)T_{1}(E)G_{0}(E).
\label{Dyson1}
\end{equation}
 
Out of the previous Dyson equation, the local electronic and magnetic properties of the adatom can be reasonably described. For instance, it leads to a MAE
of -2.8 meV for the Fe adatom. To grasp the  
effect of the rest of Pt atoms, i.e. the hundreds outer Pt atoms, on the MAE, we solve a second Dyson equation to obtain the new Green function, $G_2$:
\begin{equation}
G_{2}(E) = G_{1}(E)+G_{1}(E)V_{2}G_{2}(E).
\label{Dyson2}
\end{equation}
 where the perturbing potential, $V_2$, describes simultaneously the change induced by the adatom on the additionally incorporated 217 Pt outer 
atoms ($V_2^{'}$) and their SOI ($V_2^{so}$). In fact, $V_2^{so} = \xi (E) \mathbf{L.S}$, with $\xi(E)$ being the 
strength of SOI. Thus, $V_{2}=\displaystyle\sum_{j}(V_{2j}^{'}+V_{2j}^{so})$ where the sum runs over all outer Pt atoms. In contrast to $T_1$, $V_2$ is 
limited to the rest of Pt atoms and 
is expected to be relatively small since the perturbation decays with the distance from the adatom, which permits the use of Taylor expansions when solving Eq.~\ref{Dyson2}. 
 
The change in the IDOS, $\delta N(E)$, due to the coupling of the adatom and its NN to the 
rest of the Pt substrate atoms is then given as: $-\frac{1}{\pi}\Im \mathrm{Tr}\, ln(1 - V_2 G_1(E))$, which for small $V_2$ can be expanded up to second order:
\begin{equation}
 \delta N(E) = \frac{1}{{2\pi}}\Im \mathrm{Tr} [2V_{2} G_{1}(E)+V_{2} G_{1}(E)V_{2} G_{1}(E)].
\end{equation} 
We express $G_1$ in terms of $G_0$ as given in Eq.~\ref{Dyson1},   drop terms leading to third and fourth-order processes (these are expected to be much smaller than the second order-processes) 
and find:
\begin{eqnarray} 
 \delta N &=& \frac{1}{{2\pi}}\Im \mathrm{Tr} [2V_{2} G_{0}+2V_{2} G_{0}T_{1}G_{0} +V_{2} G_{0}V_{2} G_{0}],
 \end{eqnarray} 
where the energy argument, $E$, was taken out for the sake of simplicity. Since $V_2$ is written in terms of non-SOI- and SOI-dependent terms, this allows to disentangle the previous expression:
\begin{eqnarray}
\delta N &=&\frac{1}{\pi}\Im \mathrm{Tr}\displaystyle\sum_{j} \Large{\{} V_{2j}^{'}G_{0}+ V_{2j}^{so}G_{0}\\ \nonumber
                     & &+  T_{1} G_{0}V_{2j}^{'} G_{0} + T_{1} G_{0}V_{2j}^{so} G_{0}\\ \nonumber
                     & &+  \frac{1}{2}\sum_{j'}(V_{2j}^{'}+V_{2j}^{so}) G_{0}(V_{2j'}^{'}+V_{2j'}^{so})G_{0}\Large{\}}.
\end{eqnarray}                      
In view of our interest in the band energies that depend on the rotation of the magnetic moment, i.e. contributing to the MAE, not all terms in Eq. 6 are relevant. 
For instance the term of first order in $V_2$ or $G_0$ contain either no spin orbit coupling or only the linear SOI term. 
Therefore they vanish when one evaluates the MAE. From the last term, only the contribution from the scattering at $V_2^{'}$ and at $V_2^{so}$ is finite. 
Since these atoms are only weakly spin-polarized, the latter term is negligible as verified numerically and therefore it is not considered in the following. The contribution to the band energy relevant for the MAE is then given by  
\begin{eqnarray}
-\frac{1}{\pi} \Im  \mathrm{Tr}\int\limits_{-\infty}^{E_{F}}dE \displaystyle\sum_{j}  \Large\{ T_{1} G_{0}V_{2j}^{'} G_{0}+ T_{1} G_{0}V_{2j}^{so} G_{0}\LARGE\},
\label{Final}
\end{eqnarray}  
which has to be evaluated at the different configurations $\perp$ and $||$ orientations of the magnetic moment in order to extract the MAE.

The first term is the most simple one. It is independent of the SOI of the outer Pt atoms and just describes a renormalization of 
the MAE of the small cluster consisting of the Fe atoms and the 3Pt atoms due to the scattering at the potentials $V_{2j}^{'}$ of 
the outer Pt atoms, which does not include the SOI of these atoms. Therefore we name this contribution the no-so-term. 
The second term, called so-term, is also important and describes the double scattering at the SOI-term of $T_1$ 
and the SOI potential $V_{2j}^{so}$ of the outer atoms. These two terms might therefore be described as non-local, since they 
connect the scattering at the SOI of the inner cluster with the scattering at the potentials of the outer atoms.
The analogy of these non-local terms with the celebrated formula from Lichtenstein et al.~\cite{LKAG} 
for the evaluation of the magnetic exchange interactions is appealing, and as 
for the magnetic interactions, we expect these two terms to oscillate and decay with the distance between the two regions.  
Instead of the magnetic part of the potential, the scattering occurs at the SOI term but the mediation is made in both cases via the Green functions. 

In order to clarify the importance of the non-local terms in the MAE, we have therefore 
recalculated the anisotropy by switching on and off the SOIs of individual outer Pt atoms, based on 
Eq.~\ref{Final}. In this way, we demonstrate how the relatively small so and no-so contributions of an outer Pt atom changes the 
MAE of the complex system containing the Fe atom, its NN and that preselected outer Pt atom, and shows Friedel-like oscillations. 
 For this analysis, the cluster thus contains an Fe adatom, its 3 NN Pt atoms and one additional single Pt atom. That Pt atom probes the non-locality of 
the MAE following Eq.~\ref{Final} by considering it along different directions and distances away from the magnetic adatom. In this investigation and 
to simplify the discussion, we do not include 
the nearest neighboring atoms of that particular additional Pt atom in our cluster. Of course these boundary conditions will affect  the final values of the non-local contributions but the general conclusions of this work are not affected. We perform two steps: (step 1) SOI is switched on within the additional Pt atom. After removing the MAE of the Fe adatom and its NN 3 Pt atoms, we obtained the sum 
of the two terms given in Eq.~\ref{Final}. Then we proceed with (step 2) and switch off SOI, getting thereby the no-so-term, with which one extracts the so-term to the 
sum in Eq.~\ref{Final}.

\begin{figure*}
 \begin{center}
   \includegraphics[angle=90,scale=0.6]{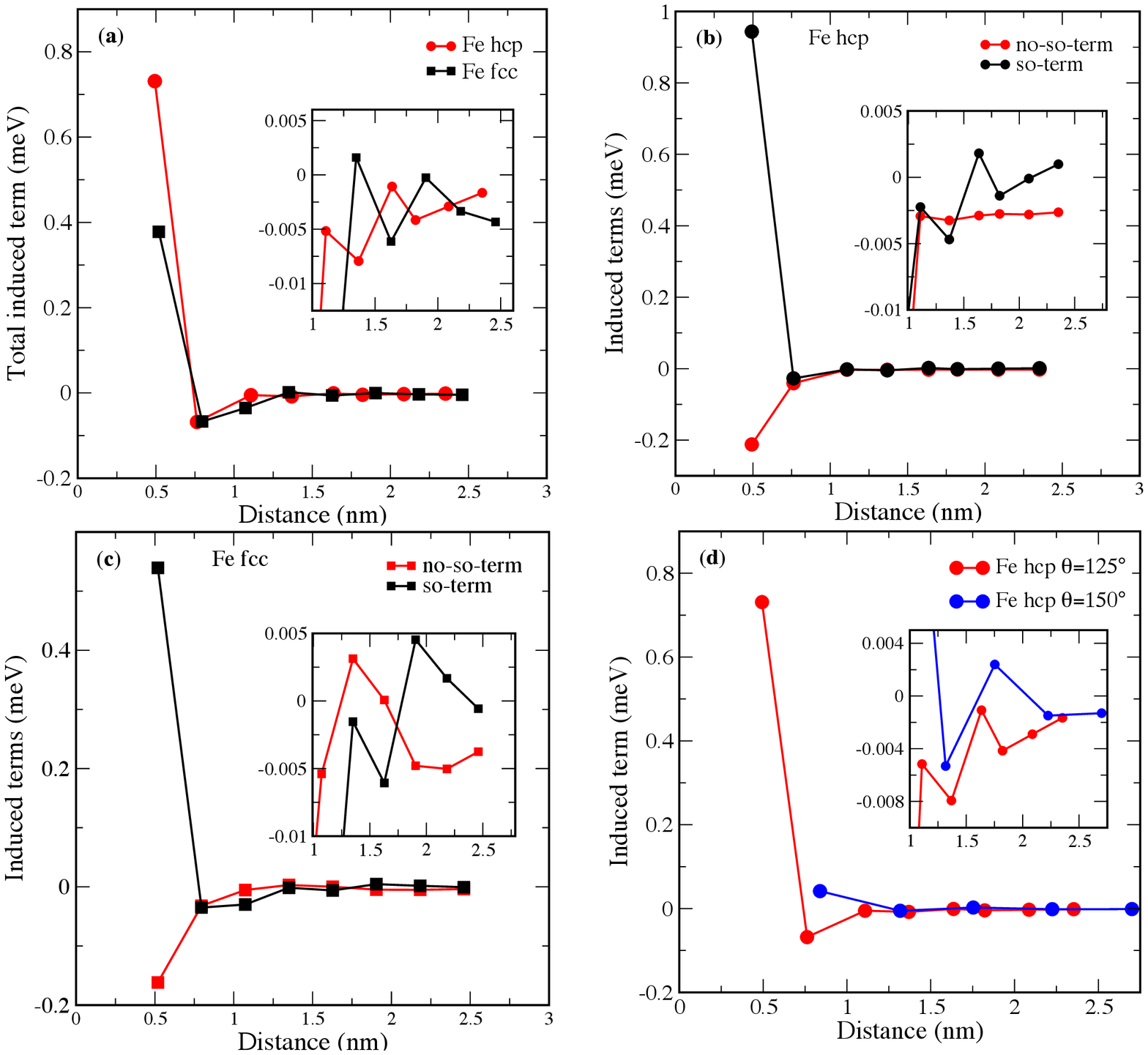}
 \end{center}
\caption{Contributions to the MAE from different shells of Pt atoms versus their distance with respect to the adatom sitting either at 
the \textit{fcc} or at the \textit{hcp} stacking sites. In (a), (b) and (c), the plotted values correspond to the Pt atoms sitting along the direction  
connecting the adatom with one of the switching Pt atoms, i.e. ($\theta=125^{\circ}, \phi=-60^{\circ}$) for the \textit{hcp} stacking site and ($\theta=128^{\circ}, \phi=30^{\circ}$) 
for the \textit{fcc} stacking site. The inset enhances the oscillations observed in the non-local terms. While in (a) the sum of the non-local contributions to the MAE is plotted, 
in (b-c)  the no-so and so-terms are plotted separately for respectively the \textit{hcp} and \textit{fcc} sites. 
(d) Anisotropy of the non-local contribution to the MAE obtained for two different set of angles: 
($\theta=125^{\circ},  \phi=-60^{\circ}$) compared to ($\theta=150^{\circ}, \phi=80^{\circ}$). The magnitude of MAE is clearly more 
enhanced along the direction passing by the switching Pt atom, i.e. the red curve. } 
\label{MAE_5}  
\end{figure*}

Fig.~\ref{MAE_5} shows the non-local contributions from a single Pt atom as 
function of  the distance, $d$, from the adatom for \textit{hcp}- 
and \textit{fcc}-sites along two directions connecting the adatom to one of the Pt switching atoms. 
While in Fig.~\ref{MAE_5}(a) we plot the sum of the non-local contributions, in Fig.~\ref{MAE_5}(b) and 
(c) these contributions are resolved into the so and no-so-terms for respectively the \textit{hcp} and \textit{fcc} sites. 
In Figs.5(a, b, and c), the chosen polar and azimuthal angles ($\theta, \phi$) are ($125^{\circ}$, $-60^{\circ}$) (\textit{hcp} stacking sites) and 
($128^{\circ}$, $30^{\circ}$) (\textit{fcc} stacking sites). 
Naturally, here we allow for an error bar for the angles ($\delta\theta=\pm3^{\circ}$ 
and $\delta\phi=\pm8^{\circ}$) since a straight line will not cross a sufficient 
number of Pt atoms at reasonable distances.
One clearly sees, that the sum of non-local terms are important outside 
the small inner region with the largest contribution emanating from the switcher atom, 
which reaches a value of $0.37$ meV for the Fe \textit{fcc}-site and $0.73$ meV for the \textit{hcp}-site.
As explained earlier, since there are only three switching atoms for the \textit{fcc}-site instead of six for 
the \textit{hcp}-site, the barrier given by the MAE of the adatom and its NN is not overcome. 
By increasing the distance from the adatom, 
the induced term oscillates and changes even sign. Its magnitude, however, is not sufficient to 
overcome the aforementioned barrier. These oscillations as function of distance have a Friedel-like character 
and are similar to those obtained for long-ranged magnetic exchange interactions~\cite{LKAG,Lounis}.

From Figs.~\ref{MAE_5}(b) and ~\ref{MAE_5}(c), we notice that the so-term is not behaving 
similarly to the no-so-term. These two terms can counteract each other as for the contribution from the switcher atom. 
Thus, for this particular atom the so-term is dominant and favors an in-plane orientation of the moment in contrast to the no-so-term. 
For large distances both terms oscillate non-trivially. Although the values plotted in Fig.~\ref{MAE_5} can look small at first sight, one should not forget that these 
are contributions from a single Pt atom. At the end, one has to sum up contributions from all the surrounding Pt atoms to get the full-non local 
part of the MAE. 

\begin{figure}[!h]
\begin{center}
\includegraphics*[angle=-90,width=0.88\linewidth]{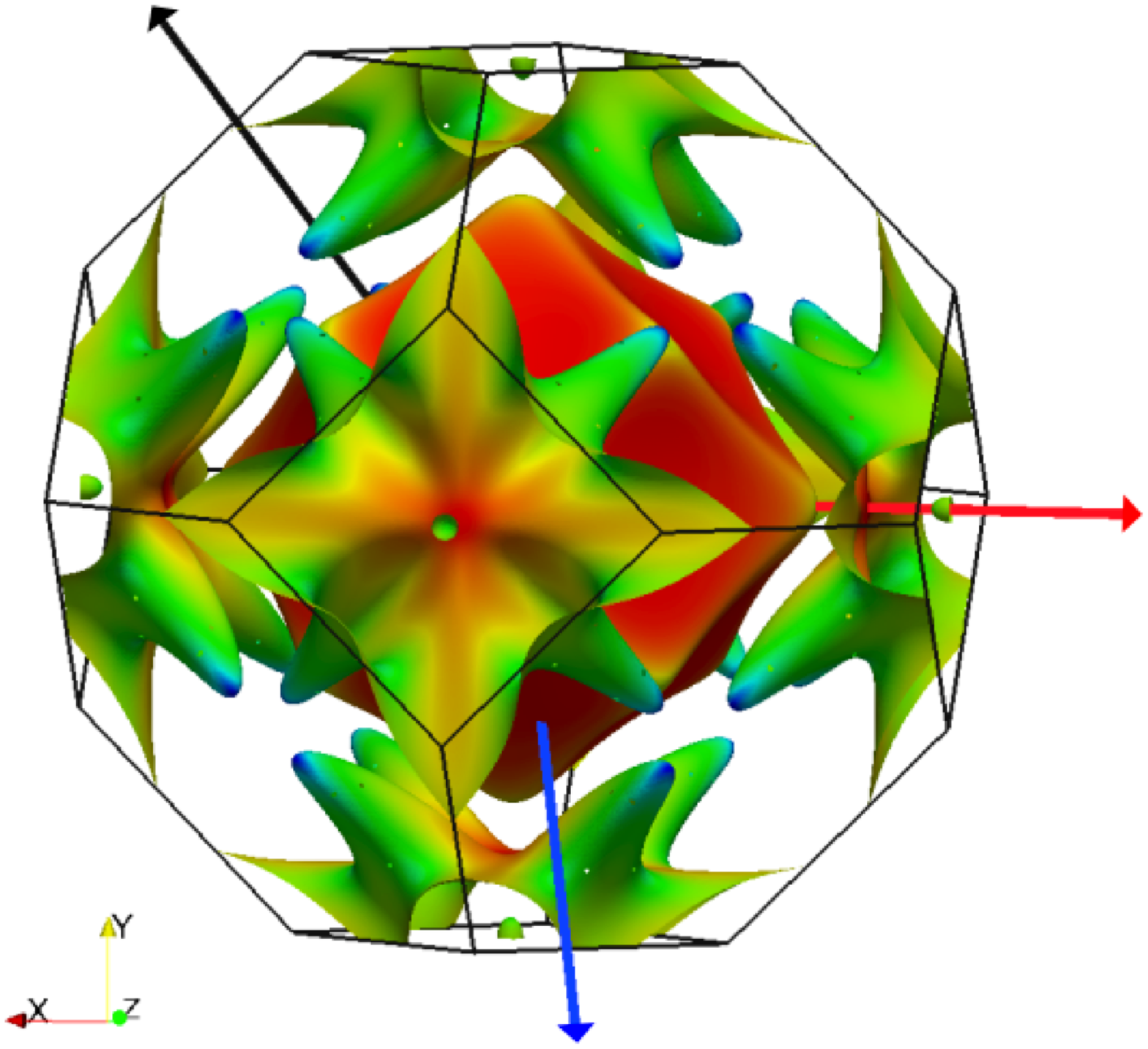}
\end{center}
\caption{Bulk Fermi surface of Pt, with directions of probed atoms as indicated by red and blue arrows and the [111] direction by a black arrow. The color code on the Fermi surface corresponds to the magnitude of the Fermi velocity (red and blue corresponding to high and low velocity, respectively).}
\label{MAE_6}
\end{figure}

These oscillating non-local parts of the MAE can be highly anisotropic as demonstrated in Fig.~\ref{MAE_5} (d), where two directions are probed. First, along the 
direction already shown in Fig.~\ref{MAE_5}(a) that connects the Fe-adatom with one switcher atom leading to a very large peak at $0.5$nm. The second probed direction does not cross 
such switcher atoms and interestingly the calculated values are considerably smaller at short distances but show similar Friedel-like oscillations at large distances. 
Thus, the non-local MAE contribution from the outer Pt-atoms show Friedel-like oscillations, but are highly anisotropic which is expected when looking at the Fermi surface 
of Pt presented in Fig.~\ref{MAE_6}. Indeed the Fermi surface, extracted utilizing the scheme described in Ref.~\cite{Zimmermann2016}, 
is extremely anisotropic such that isotropic oscillations resulting from 
a simple spherical Fermi surface cannot be expected in our particular system.

\subsection{Case of Cr, Mn and Co adatoms on {\it fcc} and {\it hcp} stacking sites}
\begin{figure}[ht!]
\begin{center}
\includegraphics*[angle=-90,width=0.9\linewidth]{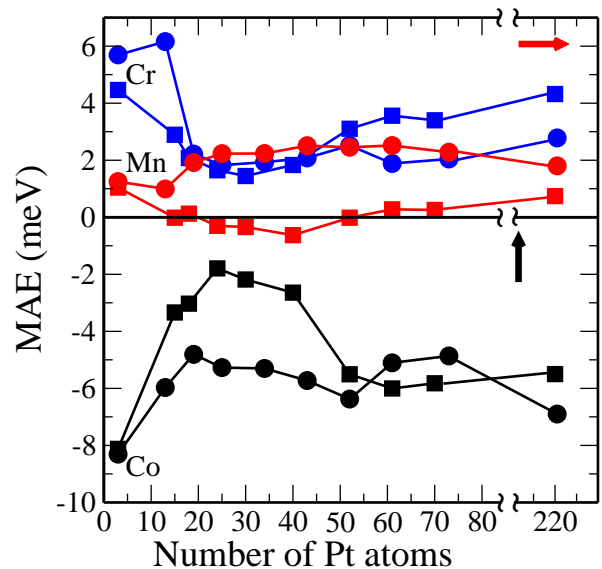}
\end{center}
\caption{MAE of Co, Mn, Cr impurities adsorbed on an \textit{fcc} (square) or on an \textit{hcp} (circles) site on top of Pt(111) surface versus the number of Pt atoms in the cluster. The convention of the sign of 
the MAE is identical to the one used in Fig.~\ref{MAE_1}.}
\label{MAE_7}
\end{figure}

For completeness, we examined the impact of the Pt spin-polarization cloud on the MAE of Cr, Mn and Co adatoms.
Like Fe adatom, Co adatom and its NN prefer an out-of-plane orientation of the magnetic moment independently from 
the binding site (Fig.~\ref{MAE_7}). The MAE found in this case ($-8.2$ meV) is however larger 
than the one of Fe adatom, making the barrier higher for an in-plane reorientation of the magnetic moment when including a large number of Pt substrate atoms (up to 221 atoms). 
Besides that, here the non-local contribution of the switching atoms to the MAE is even smaller than for Fe-adatom. The total MAE for the largest studied system decreases to 
 $-6.9$ meV and $-5.5$ meV for respectively the \textit{hcp}- and \textit{fcc}-sites. We point out that the experimental 
value of Gambardella et al.~\cite{Gambardella} is around $-9$ meV. This large value has generated a lot of theoretical investigations based on density functional theory. 
Usual simple exchange and correlation functionals, such as the local spin density approximation (LDA) or the 
generalized gradient approximation (GGA) lead to rather small MAE. Therefore, correlation effects beyond LDA or GGA were considered, e.g. by including a correlation 
$U$ as a parameter or the orbital polarization scheme to tune the MAE and understand the origin of its large magnitude. Our work demonstrates that even without the invoked 
correlation effects, the non-local contribution to 
the MAE, not considered up to now, can be crucial in the case of Co as well. We predict that in the case of the \textit{hcp}-stacking site the MAE reaches $\sim-7$ meV.
 
The case of Mn is interesting since contrary to what has been observed for Fe and Co, both the local and non-local contributions to the MAE from the switching Pt atom favor 
an in-plane orientation of the magnetic moment. However, the rest of Pt atoms are decisive. By increasing their number, 
 the adatom on the  \textit{fcc} binding-site switches first to an out-of-plane magnetic orientation before 
converging to an in-plane orientation. Cr adatom behaves similarly to Mn, i.e. both the local and non-local contributions to the MAE favor an in-plane orientation 
of the magnetic moment but unlike Mn, the local term is large: $+5.6$ meV and $+4.5$ meV for respectively the \textit{hcp} and \textit{fcc} stacking sites.
Furthermore, when compared to Mn, Fe and Co adatoms, the switching atoms at the vicinity of Cr adatom contribute to the MAE differently and favor  an out-of-plane 
orientation of the moment. This contribution is, however, not large enough to overcome 
the barrier created by the adatom and its NN. When the rest of Pt atoms are included, Cr-adatoms on both binding sites prefer an in-plane magnetic orientation.

{By changing the chemical nature of the adatom, the non-local behavior of the MAE is modified. As it can be realized from Eq.~\ref{Final}, the scattering properties 
at the adatom site, described by $T_1$, can renormalize strongly the total MAE. $T_1$ depends obviously on the electronic properties of the adatom and its nearest surrounding. 
It is not a single number but a matrix and therefore after taking the trace in 
Eq.\ref{Final} besides the impact on the magnitude of the MAE, non-trivial interference effects can occur, which affect the oscillating behavior of the MAE.}

\section{Discussions and conclusions}
To summarize, for 3{\textit{d}} adatoms on Pt(111) we demonstrated the existence of long-range, RKKY-like, 
contributions to the MAE mediated by the electronic states of the substrate. 
Since they oscillate as a function of the distance with different kind of decaying factors, they affect the
 magnitude of the total MAE and can even switch its sign. 
This depends on the details of the electronic structure, and as for Friedel oscillations or RKKY interactions, 
they can be highly anisotropic with a possibility of observing a focusing effect induced by 
the shape of the constant energy contours (e.g. the Fermi surface)~\cite{Weismann,Lounis,Bouhassoune}. Our results go beyond the approximations 
assumed along our theoretical investigations. We expect non-negligible non-local contributions to the MAE independently from 
the assumptions related to the exchange and correlation functionals, geometrical relaxations, and inclusion of a $U$ as done in traditional LDA + $U$.

The 
established effect is expected to occur in other substrates with high polarizability (e.g. Rh, W, Ir, Pd substrates), but also when confined 
electronic states are present in 
low dimensional systems (e.g. surface states of Ag and Au(111) surfaces) since the latter 
favor a lower decay of the usual Friedel oscillations. We believe that such an effect is active in the recently investigated surfaces of 
CuN/Cu(001)~\cite{Hirjibehedin,Otte} and Graphene/Rh(111)~\cite{Ternes} were unusual behavior of the MAE of different types of adsorbates 
has been observed. To verify experimentally the theoretical facts described in our work, one would have for example to switch off/on the spin-orbit interaction of a remote substrate 
Pt atom at will. This is certainly impossible, however, we believe that the signature of the non-locality of the MAE could be detectable for two magnetic adatoms on a surface, 
for example two Fe adatoms on a Pt(111) surface. We expect the MAE to be dependent on the inter-adatom distances, which is expect to be related to the non-local effect discussed 
in the main text. Thus, we expect and oscillatory behavior of the MAE measurable with state-of-the-art inelastic scanning tunneling spectroscopy, wherein the MAE leads to 
a gap in the excitation spectra.

We acknowledge fruitful discussions with Stefan Bl\"ugel and the teams of Jens Wiebe and Alex Khajetoorians. 
This work is supported by the HGF-YIG Programme VH-NG-717 (Functional Nanoscale Structure and Probe Simulation Laboratory--Funsilab) 
and the DFG project LO 1659/5-1.


\begin{thebibliography}{0}%
\makeatletter
\providecommand \@ifxundefined [1]{%
 \@ifx{#1\undefined}
}%
\providecommand \@ifnum [1]{%
 \ifnum #1\expandafter \@firstoftwo
 \else \expandafter \@secondoftwo
 \fi
}%
\providecommand \@ifx [1]{%
 \ifx #1\expandafter \@firstoftwo
 \else \expandafter \@secondoftwo
 \fi
}%
\providecommand \natexlab [1]{#1}%
\providecommand \enquote  [1]{``#1''}%
\providecommand \bibnamefont  [1]{#1}%
\providecommand \bibfnamefont [1]{#1}%
\providecommand \citenamefont [1]{#1}%
\providecommand \href@noop [0]{\@secondoftwo}%
\providecommand \href [0]{\begingroup \@sanitize@url \@href}%
\providecommand \@href[1]{\@@startlink{#1}\@@href}%
\providecommand \@@href[1]{\endgroup#1\@@endlink}%
\providecommand \@sanitize@url [0]{\catcode `\\12\catcode `\$12\catcode
  `\&12\catcode `\#12\catcode `\^12\catcode `\_12\catcode `\%12\relax}%
\providecommand \@@startlink[1]{}%
\providecommand \@@endlink[0]{}%
\providecommand \url  [0]{\begingroup\@sanitize@url \@url }%
\providecommand \@url [1]{\endgroup\@href {#1}{\urlprefix }}%
\providecommand \urlprefix  [0]{URL }%
\providecommand \Eprint [0]{\href }%
\providecommand \doibase [0]{http://dx.doi.org/}%
\providecommand \selectlanguage [0]{\@gobble}%
\providecommand \bibinfo  [0]{\@secondoftwo}%
\providecommand \bibfield  [0]{\@secondoftwo}%
\providecommand \translation [1]{[#1]}%
\providecommand \BibitemOpen [0]{}%
\providecommand \bibitemStop [0]{}%
\providecommand \bibitemNoStop [0]{.\EOS\space}%
\providecommand \EOS [0]{\spacefactor3000\relax}%
\providecommand \BibitemShut  [1]{\csname bibitem#1\endcsname}%
\let\auto@bib@innerbib\@empty
\end{thebibliography}%


\begin{thebibliography}{99}
\bibitem{Gambardella}{P. Gambardella, S. Rusponi, M. Veronese, S. S. Dhesi, C. Grazioli, A. Dallmeyer, I. Cabria, R. Zeller, 
P. H. Dederichs, K. Kern, C. Carbone, and H. Brune, Science {\bf{300}}, 1130 (2003)}
\bibitem{Bode} {M. Bode, O. Pietzsch, A. Kubetzka, R. Wiesendanger, Phys. Rev. Lett. {\bf{92}}, 067201 (2004)}
\bibitem{Rau}{I. G. Rau, S. Baumann, S. Rusponi, F. Donati, S. Stepanow, L. Gragnaniello, J. Dreiser, C. Piamonteze, 
F. Nolting, S. Gangopadhyay, O. R. Albertini, R. M. Macfarlane, C. P. Lutz, B. A. Jones, P. Gambardella, A. J. Heinrich, H. Brune, 
Science {\bf 344}, 988 (2014)}
\bibitem{Khajetoorians_PRL}{A. A. Khajetoorians, S. Lounis, B. Chilia, A. T. Costa, L. Zhou, D. L. Mills, J. Wiebe, R. Wiesendanger, 
Phys. Rev. Lett. {\bf 106}, 037205 (2011)}
\bibitem{Dubout}{Q. Dubout, F. Donati, C. W\"ackerlin, F. Calleja, M. Etzkorn, A. Lehnert, L. Claude, 
P. Gambardella, H. Brune, Phys. Rev. Lett. {\bf 114}, 106807 (2015)}
\bibitem{Honolka}{J. Honolka, T. Y. Lee, K. Kuhne, A. Enders, R. Skomski, S. Bornemann, S. Mankovsky, J. Minar, J. Staunton, H. Ebert, M. Hessler, K. Fauth, G. Sch\"utz, A.Buchsbaum, M. Schmid, P. Varga, and K. Kern, Phys. Rev. Lett. {\bf 102}, 067207 (2009)}
\bibitem{Khajetoorians_Science}{A. A. Khajetoorians, B. Baxevanis, C. H\"ubner, T. Schlenk, S. Krause, T. O. Wehling, S. Lounis, 
A. Lichtenstein, D. Pfannkuche, J. Wiebe, R. Wiesendanger, Science {\bf 339}, 55 (2013)}
\bibitem{Otte}{B. Bryant, A. Spinelli, J. J. T. Wagenaar, M. Gerrits, A. F. Otte, Phys. Rev. Lett. {\bf 111}, 127203 (2013)}
\bibitem{Krause}{S. Krause, L. Berbil-Bautista, G. Herzog, M. Bode, R. Wiesendanger, Science {\bf 317}, 1537 (2007)}
\bibitem{Gambardella2}{P. Gambardella, A. Dallmeyer, K. Maiti, M. C. Malagoli, W. Eberhardt, K. Kern, C. Carbone, Nature {\bf 416}, 
301 (2002)}
\bibitem{Sessoli} {R. Sessoli, D. Gatteschi, A. Caneschi, M. A. Novak, Nature {\bf{365}}, 141 (1993)}
\bibitem{Gatteschi} {D. Gatteschi, R. Sessoli, J. Villain, Molecular Nanomagnets, (Oxford Univ. Press, Oxford, (2006))}
\bibitem{Brede}{J. Brede, N. Atodiresei, V. Caciuc, M. Bazarnik, A. Al-Zubi, S. Bl\"ugel, R. Wiesendanger, 
Nature Nanotechnology {\bf 9}, 1018 (2014)}
\bibitem{Lodi_Rizzini}{A. Lodi Rizzini, C. Krull, T. Balashov, J. J. Kavich, A. Mugarza, P. S. Miedema, P. K. Thakur, V. Sessi, 
S. Klyatskaya, M. Ruben, S. Stepanow, P. Gambardella, Phys. Rev. Lett. {\bf 107} (2011)}
\bibitem{Khajetoorians}{A. A. Khajetoorians, T. Schlenk, B. Schweflinghaus, M. dos Santos Dias, M. Steinbrecher, M. Bouhassoune, S. Lounis, J. Wiebe, 
R. Wiesendanger, Phys. Rev. Lett. {\bf{111}}, 157204 (2013)}
\bibitem{Sipr}{O. Sipr, S. Bornemann, J, Minar, and H. Ebert, Phys. Rev. B {\bf{82}}, 174414 (2010)}
\bibitem{Meier}{F. Meier, S. Lounis, J. Wiebe, L. Zhou, S. Heers, Ph. Mavropoulos, P. H. Dederichs, S. Bl\"ugel, R. Wiesendanger, 
Phys. Rev. B {\bf 83}, 075407 (2011)}
\bibitem{Nieuwenhuys}{G. J. Nieuwenhuys, Adv. Phys. {\bf 24}, 515 (1975)}
\bibitem{Herrmannsdorfer}{T. Herrmannsd\"orfer, S. Rehmann, W. Wendler, F. Pobell, J. Low Temp. Phys. {\bf 104}, 49 (1996)}
\bibitem{Oswald}{A. Oswald, R. Zeller, P. H. Dederichs, Phys. Rev. Lett. {\bf 56}, 1419 (1986)}
\bibitem{Swieca}{K. Swieca, Y. Kondo, and F. Pobell, Phys. Rev. B {\bf 56}, 6066 (1997)}
\bibitem{Mitani}{S. Mitani, K. Takanashi, M. Sano, H. Fujimori, A. Osawa, and H. Nakajima, J. Magn. Magn. Mater. {\bf 148}, 163 (1995)}
\bibitem{Blonski}{P. Blonski, A. Lehnert, S. Dennler, S. Rusponi, M. Etzkorn, G. Moulas, P. Bencok, P. Gambardella, H. Brune, and J. Hafner, Phys. Rev. B {\bf81}, 104426 (2010)}
\bibitem{Szunyogh}{L. Szunyogh, and B. L. Gyorffy, Phys. Rev. Lett. {\bf{78}}, 3765 (1997)}
\bibitem{Szunyogh2}{L. Szunyogh, G. Zarand, S. Gallego, M. C. Munoz,  and B. L. Gyorffy, Phys. Rev. Lett. {\bf{96}}, 067204 (2006)}
\bibitem{Aas} {C. J. Aas, K. Palot\'as, L. Szunyogh, and R. W. Chantrell, J. Phys.: Condens. Matter {\bf 24} 406001 (2012)} 
\bibitem{Sigalas} {M. M. Sigalas and D. A. Papaconstantopoulos, Phys. Rev. B {\bf 50}, 7255 (1994)}
\bibitem{Friedel}{J. Friedel, Nuovo Cim. {\bf 7} (suppl. 2), 287 (1958)} 
\bibitem{RKKY}{M. A. Ruderman, C. Kittel, Phys. Rev. {\bf 96}, 99 (1954); T. Kasuya, Prog. Theor. Phys. {\bf 16}, 45 (1956); 
K. Yosida, Phys. Rev. {\bf 106}, 893 (1957)}
\bibitem{Crommie1}{Crommie, M. F., Lutz, C. P., \& Eigler, D. M. Imaging standing waves in a two-dimensional electron gas. Nature \textbf{363}, 524-527 (1993)} 
\bibitem{Avouris1}{Hasegawa, Y., \& Avouris, Ph. Direct Observation of standing wave formation at surface steps using scanning tunneling spectroscopy. Phys. Rev. Lett. \textbf{71}, 1071-1074 (1993)}
\bibitem{Zhou}{L. Zhou, J. Wiebe, S. Lounis, E. Vedmedenko, F. Meier, P. H. Dederichs, S. Bl\"ugel, and R. Wiesendanger, Nat. Phys. {\bf 6}, 187 (2010)}
\bibitem{Khajetoorians2}{A. A. Khajetoorians, J. Wiebe, B. Chilian, S. Lounis, S. Bl\"ugel, and R. Wiesendanger, Nat. Phys. {\bf 8}, 497 (2012)}
\bibitem{Weismann}{A. Weismann, M. Wenderoth, S. Lounis, P. Zahn, N. Quaas, R. G. Ulbrich, P. H. Dederichs, and S. Bl\"ugel, Science {\bf{323}}, 1190 (2009)}
\bibitem{Lounis}{ S. Lounis, P. Zahn, A. Weismann, M. Wenderoth, R. G. Ulbrich, I. Mertig, P. H. Dederichs, and S. Bl\"ugel, Phys. Rev. B {\bf{83}}, 035427 (2011)}
\bibitem{Avotina}{Ye. S. Avotina, Yu. A. Kolesnichenko, A. N. Omelyanchouk, A. F. Otte, and J. M. van Ruitenbeek, Phys. Rev. B {\bf 71}, 115430 (2005)}
\bibitem{Bouhassoune}{M. Bouhassoune, B. Zimmermann, Ph. Mavropoulos, D. Wortmann, P. H. Dederichs, S. Bl\"ugel, S. Lounis, Nature Communications {\bf 5}, 5558 (2014)}
\bibitem{Pruser}{H. Pr\"user, P. E. Dargel, M. Bouhassoune, R. G. Ulbrich, T. Pruschke, S. Lounis, M. Wenderoth, Nature Communications {\bf 5}, 5417 (2014)}
\bibitem{Lounis_PRL}{Lounis, S., Bringer, A., Bl\"ugel, S. Magnetic adatom induced skyrmion-like spin texture in surface electron waves. Phys. Rev. Lett. {\bf 108}, 207202 (2012)}
\bibitem{Mackintosh}{A. R. Mackintosh and O. K. Andersen, in Electron at the Fermi Surface, edited by M. Springford (Cambridge University Press, Cambridge, England, 1980), p. 149}
\bibitem{Jansen}{H. J. F. Jansen, Phys. Rev. B {\bf59}, 4699 (1999)}
\bibitem{Papanikolaou}{N. Papanikolaou et al. J. Phys. Condens. Matter {\bf14}, 2799 (2002)}
\bibitem{Bauer}{D. Bauer, PhD Thesis, Forschungszentrum J\"ulich and RWTH Aachen (2014)}
\bibitem{LDA}{S. Vosko, L. Wilk, and M. Nusair, Can. J. Phys. {\bf 58}, 1200 (1980)}
\bibitem{Blonski2}{P. Blonski and J. Hafner, J. Phys.: Condens. Matter {\bf 21}, 426001 (2009)} 
\bibitem{Schweflinghaus2016}{B. Schweflinghaus, M. dos Santos Dias, S. Lounis, Phys. Rev. B {\bf 93}, 035451 (2016)}
\bibitem{Lloyd} {P. Lloyd, Proc. Phys. Soc. {\bf 90}, 207 (1967)}
\bibitem{LKAG}{A. I. Liechtenstein, M. I. Katsnelson, V. P. Antropov, V. A. Gubanov, J. Magn. Magn. Mater. {\bf 67}, 65 (1987)}
\bibitem{Lounis2}{S. Lounis, P. H. Dederichs, Phys. Rev. B (R) 82, 180404 (2010)}
\bibitem{Zimmermann2016} {B. Zimmermann, Ph. Mavropoulos, N. H. Long, C.-R. Gerhorst, S. Bl\"ugel, Y. Mokrousov, Phys. Rev. B {\bf 93}, 144403 (2016)}
\bibitem{Hirjibehedin}{J. C. Oberg, M. R. Calvo, F. Delgado, M. Moro-Lagares, D. Serrate, D. Jacob, J. Fernandez-Rossier, C. F. Hirjibehedin, 
Nature Nano. {\bf 9}, 64 (2014)}
\bibitem{Ternes}{P. Jacobson, T. Herden, M. Muenks, G. Laskin, O. Brovko, V. Stepanyuk, M. Ternes, K. Kern, Nature Comm. {\bf 6}, 8536 (2015).}
\end{thebibliography}
\end{document}